\documentclass[12pt,preprint]{aastex}

\begin{document}

\title{Giant eruptions of very massive stars\altaffilmark{1}  }  
\author{ Kris Davidson\altaffilmark{2} }  
\altaffiltext{1} {To appear in {\it Physics \& Chemistry of Late 
      Stages of Stellar Evolution,\/} 
      11th Pacific Rim Astrophysics Conference, Hong Kong December 2015 
      (ed.\ S.\ Kwok \& K.C.\ Leung, 2016) } 
      
\altaffiltext{2}{School of Physics and Astronomy, University of Minnesota, 
    Minneapolis MN 55455, U.S.A. (kd@astro.umn.edu)  } 

\begin{abstract}
Giant eruptions or supernova-impostor events are far more mysterious than 
true supernovae.  An extreme example can release as much radiative energy 
as a SN, ejecting several $M_\odot$ of material.  These events involve  
continuous radiation-driven outflows rather than blast waves.  They 
constitute one of the main  unsolved problems in stellar astrophysics,  
but have received surprisingly little theoretical effort.  Here I note 
some aspects that are not yet familiar to most astronomers.  
\end{abstract}

\section{Introduction}  


Let me steal a metaphor from Tom Wolfe.  Some of us think that a Demon 
lives near the Eddington Limit.  If you (as a massive star) try to 
approach that limit, he intervenes before you get very close to it. 
He shakes you so violently that you lose mass and energy, and throws 
you back away from the edge.  After watching this happen to a number 
of stars, we have never seen the Demon's face.  In hindsight his 
behavior almost makes sense in terms of physics, and it dramatically 
alters the evolution of very massive stars.  But the only certain 
factor is that no theorist predicted it.  


This idea grew from several disparate topics.  Three decades ago, 
Luminous Blue Variable stars (LBVs) attracted attention because 
their sporadic mass-loss events could explain  why there are no yellow 
and red supergiants above $L  \sim  \, 10^{5.9} \, L_\odot$ [1].   
LBVs are closer to $(L/M)_{\mathrm {Edd}}$ than other stars in the same part 
of the H-R diagram.  Meanwhile the parameters of $\eta$ Carinae's huge 
outburst in 1830--1860 became clear;   as usual one really good 
example provided better clues than dozens of less extreme ones.  
That event expelled 10 to 30 $M_\odot$ of material and the 
same amount of light as a typical supernova, in a timescale roughly 
100 times as long as a SN event, and the star survived.  
By 1990 the role of episodic mass 
loss in the most massive stars was widely recognized [2].   
But then an odd thing happened: After 1995, astronomers almost seemed 
to forget this topic!  When mass loss in ordinary hot-star winds was 
reassessed downward around 2005, invalidating the published 
evolutionary tracks [3], experts began to propose eruptive loss instead.  
In fact its role had been familiar to many people 15 years earlier. 


Meanwhile, SN surveys revealed abnormal explosions in other 
galaxies.  Instead of obvious blast waves, they produced slower,  
more continuous outflows resembling $\eta$ Car's great 
eruption [4].  Some were labeled 
``Type IIn supernovae,'' which implies pre-existing circumstellar 
ejecta.  And occasionally the star survived!  We call such cases  
Supernova Impostors, with $\eta$ Car as the obvious prototype.   
Equally embarrassing, a few SN events had precursor outbursts,  
which seemed paradoxical in any textbook view.   
One supernova observed in 2012 already had the name SN 2009ip!    


Few researchers believe that any one instability mechanism produced all 
the eruptions that I've mentioned.   Several types probably co-exist: 
(1) Core-collapse SNae that had unusual extended envelopes when 
they exploded;  (2) other core phenomena, mostly but not entirely 
related to nuclear processes; (3) the Demon mentioned earlier, a 
hypothetical radiative/fluid instability (or instabilities) that can 
arise when $L/M \, >  \, 0.5 \, (L/M)_{\mathrm {Edd}}$;    
(4) binary interactions, mass transfer, and/or mergers; and (5) whatever  
we haven't thought of.  Incidentally, number 4 is frequently offered as 
a panacea, but the statistics make that very unlikely.  Since 
all these conjectural types of eruption share the same radiative outflow 
physics,  {\it they look alike\/} when viewed from outside. 


As you might guess, the situation has become too confused for a newcomer 
to learn easily.  Some basic fallacies have propagated in the 
literature.  For example, most supernova specialists assume that 
progenitors of type IIn must be LBVs, based on faulty logic.  Many 
recent  authors have applied the term LBV to stars that {\it might\/} 
belong to that class but are unproven, and to other stars 
that {\it don't\/} belong.  The big evolutionary differences between 
luminous and less-luminous LBVs are seldom acknowledged.  Various 
explain-all ``models'' consist mainly of words, cartoon sketches, and/or 
computer runs with many unadvertised assumptions.  Reciting their 
defects would fill many pages.  Thus, I earnestly advise everyone to 
be wary of groupthink -- and equally wary of claims that some   
paper has revolutionized the topic. 


Since this is an account of some concepts, and not of the literature, 
few papers will be cited here.  For general background see articles 
by many authors in [5], and older references in [2] 
-- a review  
which, amazingly, has not been superseded nor seriously disproven  
after 20 years (except that it said too little about rotation).
Each citation below implies ``and other references noted therein.''

\section{Radiative physics in a giant eruption}  

Let's begin with a clear definition.  A stellar ``giant eruption'' is 
a {\it super-Eddington mass outflow,\/} driven by continuum 
radiation pressure.  It is not driven by a shock 
wave, though shocks may propagate through the flow.  It's  
opaque, so the  photosphere is located at a fairly 
large radius in the outflow.  The eruption usually persists for 
months or years, much longer than any relevant dynamical timescale.   
It is quite likely to be non-spherical, e.g.\ the famously bipolar  
case of $\eta$ Car.  The word ``eruption'' is especially apt in 
some models that behave like geysers, with instability propagating 
inward while expelling mass outward [2].    
Ordinary large LBV outbursts are not giant eruptions, but they have 
physical similarities. 


We often mention the Eddington factor $\Gamma \, = \, L/L_{\mathrm {Edd}}$, 
where  $L_{\mathrm {Edd}} \, \approx \, 4{\pi}cGM/{\kappa}$.   
Most of the opacity $\kappa$ is due to Thomson scattering, 
but the relatively small absorption opacity determines the size of 
the photosphere (see below).  Eta Car's great 
eruption had $\Gamma \, \sim \, 2$ to $10$, but SN 2011ht and SN 2009ip 
had $\Gamma \, >  \, 50$.  Occasionally it is claimed that such large 
values cast doubt on the entire concept. 
In fact, however, the basic outflow math for 
$\Gamma \, \rightarrow \, \infty$   is not much different from 
$\Gamma \, \sim \, 4$.  The star's mass $M$ then has little 
effect and everything depends on $L$ and on the sonic point 
where the flow originates;  since higher eruption luminosities 
generally have larger size scales, the outflow speeds remain   
below 1000 km s$^{-1}$ in most cases even with $\Gamma \, > \, 50$.  


In principle a giant eruption can originate in more than one way.  
The Demon instability mentioned above, for example, might be an 
``opacity-modified Eddington limit'' affair not far below 
the star's photosphere, or (more likely) it may involve strange-mode 
instabilities in the notorious high-opacity layers where 
$T \, \sim \, 3 \times 10^5$ K [2].  Either way, a lot of extra  
radiation pushes material while diffusing outward. 
A core-collapse SN can also become a giant eruption.  Initially,  
of course, a SN blast wave occurs.   In a normal 
case it eventually reaches the star's surface, with familiar results. 
But suppose the star is surrounded by a large opaque envelope 100 times 
as dense as an ordinary stellar wind -- something resembling the 
wind that $\eta$ Car had a century ago.   It's easy to show that 
photons then diffuse outward ahead of the blast wave.  
(See, e.g., [9].  Supernova enthusiasts perversely call this 
phenomenon ``shock breakout,''  but it's really photon breakout.) 
The diffusing radiation accelerates a giant eruption that precedes 
the blast wave.  The shock doesn't reach the outflow's 
photosphere until a time well after the maximum brightness.   


Logically, this account of a giant-eruption SN transfers the 
problem to {\it why\/} that dense circumstellar stuff was there.  
It requires a big precursor outflow, a less luminous giant eruption 
in the last few years before the main explosion. 
But this seems counter-intuitive, 
because the tiny pre-SN core with its rapid nuclear timescale is not 
supposed to know about the star's outer layers, and vice-versa.  
This looks like evidence that the Demon lives in the core of 
the star, rather than the outer layers as some of us have usually 
supposed.  But if that's true, then why is there an LBV instability 
strip in the HR diagram, representing only the outer layers?  
And why does it explain the HRD's upper limit so nicely? 
(See [2]).  Are there two different Demons?  Or more?    
These are among the biggest questions in stellar astrophysics, 
because no one has a credible answer yet.  


Unfortunately a super-Eddington flow is difficult to calculate, 
because 3-dimensional effects may be crucial.  
The ejected mass and velocities in $\eta$ Car's giant eruption  
do not match simple 1-D outflow calculations [6,7].  This 
shouldn't surprise us, since it's conceptually ``a light fluid 
driving a heavy one'' \`{a} la Rayleigh-Taylor.  R-T instabilities 
within a star imply convection, but a giant eruption is a supersonic 
outflow.  Likely result:  local mass concentrations form and photons 
escape preferentially along the easiest paths between blobs -- 
thus reducing the effective $\kappa$ so far as radiative acceleration 
is concerned.  Indeed the ejecta around $\eta$ Car show obvious granulation 
with reasonable size scales.  This phenomenon in a super-Eddington 
flow has been called ``porosity'' [7], though ``granulation'' may 
be a better term depending on the topology of the mass condensations.   
In order to avoid having to do a fresh 3-D model for every 
observed eruption, {\it we need a general, albeit rough, empirical 
prescription\/} based on many numerical simulations -- in 
the same spirit as mixing length theory for convection. 
(Some authors have recently asserted that 1-D models work better 
than I said above, see [8] and useful refs.\ therein;  but if this 
is true, it needs to be confirmed by 3-D investigations.) 


Next let me say something about observed continuum slopes and 
emission lines.  Giant eruptions, LBV 
eruptions, and other mass outflows typically have apparent temperatures 
between 7000 and 9000 K at maximum brightness [2]. 
This fact is a consequence of opacity physics, and does not imply 
that the outbursts had similar causes.  The average temperature  
of escaping radiation represents the ``thermalization depth'' where 
$\surd \, (3 \, {\tau}_{\mathrm tot} {\tau}_{\mathrm abs}) \, \sim \, 1$.  
A crucial fact is that opacity decreases rapidly below $T \sim$ 7500 K, 
and {\it very\/} rapidly below 6500 K.  Thus we find some interesting  
generalities for a wind or outflow with a given luminosity [10]. 
First, a moderate mass-loss rate can produce apparent temperatures 
around 8000 K, defined in a particular way. But reducing that to 
7000 K requires a substantially larger density or mass-flow rate;  
6000 K implies a rather huge rate;  and much lower temperatures 
are unlikely in practical terms.  (This is analogous to 
the Hayashi limit.)  Since temperatures in this range also have 
bolometric corrections near zero,  it is entirely natural that opaque 
outflows of all kinds often look like $T \sim 7500$ K at maximum brightness. 


({\it Caveat:}  These temperatures are defined in a 
particular way [10], and other definitions give different values. 
The temperature at $\tau \, \approx \, 2/3$ is {\it not\/} a good 
choice, since it has no physical significance in a convex diffuse  
configuration.)


And here's  a nice consequence for emission lines.  The 
thermalization depth mentioned above usually occurs at Thomson 
scattering depths ${\tau}_{\mathrm sc} \, \sim$ 2 or 3,  
a consequence of the opacity dependences.  Emission lines are 
formed in the diffuse outflow outside that radius, i.e.\ in 
regions where ${\tau}_{\mathrm sc} \, \sim$ 1 to 2.  This is the 
range where Thomson-scattered line wings are apparent, 
recognizable, and moderate -- just like  the spectra of 
the best-observed giant eruptions (e.g.\ [11,12]). 
{\it When we see moderate Thomson-scattering wings on the 
Balmer emission lines, with a visual continuum slope like  
$T \sim$ 7000 to 12000 K, then we're probably looking at a super-Eddington 
flow.} 


At this point I feel bound to warn against a particular spectroscopic 
fallacy that has caused confusion.  {\it Absorption-line spectra of opaque 
winds cannot be classified with stellar spectral types.\/} 
Compare, for instance, a star with $T_{\mathrm eff} = 6500$ K vs.\ an  
opaque wind with the same photosphere temperature.  The star's atmosphere 
has practically no material below 5200 K, but outer parts of the wind 
can be substantially cooler than that.  Therefore it is possible for 
the wind to show ``cool'' absorption features along with those that 
we expect to see in a 6500 K spectrum.  This pitfall led to 
a serious misinterpretation of the light-echo spectrum of $\eta$ Car's 
great eruption. Contrary to some well-publicized claims [13], in fact 
that spectrum seems reasonably consistent with the super-Eddington flow 
type of model [14].  It would be helpful, though, to have some genuine 
theoretical spectrum models for this case -- a difficult undertaking. 


We now have a sizable fund of excellent data on LBVs and related stars, 
$\eta$ Car and other supernova impostors, and on giant-eruption  
supernovae;  observers have done their job well.  But theorists
have given this topic far less attention than it deserves.  As 
I implied earlier, {\it this subject is relatively unexplored 
territory\/} for theory.    
   
   \vspace{2mm}   
   \hspace{65mm}  * * * 

Finally here's a second gratuitous literary allusion.  So far as I know, 
ref.\ [15] contains the earliest account of a ``continuing explosion'' 
qualitatively reminiscent of a stellar eruption.  It employed 
Carolinum, and $\eta$ Car used to be in a constellation named Robur 
Carolinum; so maybe that author knew something that we don't.   On the 
other hand, he favored a nuclear process that is difficult to scale up 
to the size of a giant eruption. 

   \vspace{8mm}

\noindent  References   
\\  ( 1) \ Humphreys, R.M. \& Davidson, K. 1979, \apj, 232, 409.
\\  ( 2) \ Humphreys, R.M. \& Davidson, K. 1994, \pasp, 106, 1025.  
\\  ( 3) \ Fullerton, A.W., Massa, D.J., \& Prinja, K. 2006, 
     \apj, 637, 1025.  
\\  ( 4)  \ Van Dyk, S.D. \& Matheson, T. 2012, in ref.\ 5, p.\ 249.  
\\  ( 5) \ Davidson, K., \& Humphreys, R.M. (eds.) 2012, 
    ASSL 384, {\it Eta Carinae and the Supernova Impostors\/}  
    (Springer, New York):  articles by various authors.  
\\  ( 6)   \ Shaviv, N. 2000, \apj, 532, L137.  
\\  ( 7)   \ Owocki, S. \& Shaviv, N. 2012, in ref.\ 5, p.\ 275. 
\\  ( 8)  \ Quataert, E., et al 2015, arXiv:1509.06370.    
\\  ( 9)  \ Margutti, R., et al 2014, \apj, 780, 21.    
\\  (10)  \ Davidson, K. 1987, \apj, 317, 760.  
\\  (11)  \ Dessart, L., et al.\ 2009, 394, 21.   
\\  (12)  \ Humphreys, R.M., et al.\ 2012, \apj, 760, 93.   
\\  (13)  \ Rest, A., et al.\ 2012, Nature, 482, 375. 
\\  (14)  \ Davidson, K. \& Humphreys, R.M. 2012, Nature, 486, E1  
\\  (15)  \ Wells, H.G. 1914, {\it The World Set Free\/} 
   (MacMillan, London).

\end{document}